\documentclass[11pt]{article}
\usepackage[T1]{fontenc}
\usepackage[a4paper,margin=2.5cm]{geometry}
\usepackage{mathptmx}

\usepackage{microtype}
\usepackage{amsmath,amssymb}
\usepackage{graphicx}
\usepackage{booktabs}
\usepackage{threeparttable}
\usepackage{tabularx}
\usepackage{authblk}
\usepackage{float}
\usepackage[font=small,labelfont=bf]{caption}
\usepackage{setspace}
\usepackage{xurl}
\usepackage[numbers,sort&compress]{natbib}
\usepackage[hidelinks]{hyperref}

\floatstyle{ruled}
\newfloat{algorithm}{tbp}{loa}
\floatname{algorithm}{Algorithm}

\title{Two-Stage Student Team Formation in Practice: Balancing
Teammate Nominations and Cohort-Level Constraints}
\author[1]{Yiwei Sun}
\author[1]{Xinru Deng}
\author[1]{Dimitrios G. Papageorgiou}
\affil[1]{School of Engineering and Materials Science,
Queen Mary University of London, London, UK\\
\textit{Corresponding author:} 
\href{mailto:yiwei.sun@qmul.ac.uk}{yiwei.sun@qmul.ac.uk}}
\date{}

\hypersetup{
  pdftitle={Two-Stage Student Team Formation in Practice: Balancing Teammate Nominations and Cohort-Level Constraints},
  pdfauthor={Yiwei Sun, Xinru Deng, Dimitrios G. Papageorgiou},
  pdfkeywords={team formation, operational research in education, heuristics, constraint programming, assignment}
}

\begin{document}

\maketitle

\begin{abstract}
\noindent
Assigning students to project teams in large cohorts must reconcile
teammate preferences, academic balance, and composition constraints.
We report a two-phase procedure deployed with 238 engineering
students: Phase~1 converts teammate nominations into fixed two- or
three-student units; Phase~2 pairs the units by simulated annealing
under balance, gender-isolation, and size penalties. Against 82
historical grouping instances and 10{,}000 random partitions per
cohort, the deployment produced no gender-isolated student,
between-team GPA spread at 26\% and 34\% of the random expectation,
and a requested teammate for 80.4\% of valid submitters. Constraint
programming, conditional on the deployed units, proves the deployed pairings within 0.37 and 0.27 points of the integerised conditional optimum. A single-stage
comparator exceeds the deployment on aggregate satisfaction and
balance only at a tenfold search budget, and rarely delivers both
nominated teammates. Protected matches are invariant by construction. The open release
provides code, verification scripts, and aggregate results to support
independent reuse, audit, and extension.
\end{abstract}



\section{Introduction}\label{sec:intro}

Allocating students to project teams is an operational
decision-support problem in university administration: each term,
module coordinators must partition large cohorts into working teams
under conflicting requirements from students and staff, with limited
time and full accountability for the result. This article reports the
design, deployment, and evaluation of a decision-support tool for
that task. The central question is practical: how can a large-course
administrator preserve student teammate nominations while producing
academically balanced teams, avoiding demographic isolation, and
maintaining a transparent, reproducible workflow?

Creating project teams at scale is an allocation problem with several
competing aims. Instructors may wish to accommodate student preferences,
distribute prior attainment, avoid demographic isolation, and retain a
useful mix of skills within each team
\citep{oakley2004turning,michaelsen2008essential,kifle_bonner_2023,
wang2024systematic}. These choices become difficult to handle manually in
cohorts of more than 100 students. Local Student--Staff Liaison Committee
records from 2021--2025 also document recurring concerns about grouping
and uneven participation; one 2024--25 record states that more than 60\% of students complained about unequal contribution to teamwork (Student--Staff Liaison Committee records, 2021--2025; available from the corresponding author).

Common formation methods make different trade-offs. Random assignment is
simple and procedurally neutral but does not use information about students
\citep{pociask2017teamformation}. Self-selection allows students to
choose teammates based on existing relationships or shared
backgrounds~\citep{fernandes2024antecedents}. This can preserve social
ties, but it can also reinforce homogeneity and leave some students
unchosen. Instructor assignment based on academic
or demographic criteria can improve balance, but students may view an
opaque allocation process as arbitrary
\citep{farland2019teamstyles}.

\subsection{Algorithmic approaches to team formation}\label{sec:related}

Algorithmic tools support instructor-configured allocation at
scale. CATME forms teams from instructor-specified criteria such as
GPA distribution and gender balance \citep{layton2010catme};
\citet{purdue_catme_gs_2019} later added ranked project preferences
via the Gale--Shapley algorithm. Team Anneal applies simulated
annealing to weighted composition constraints, including isolation
rules \citep{uq_teamanneal}, and LIFT lets students propose,
vote on, and weight the formation criteria themselves
\citep{hastings2020lift}; instructors still shoulder most
configuration decisions in current practice
\citep{hastings2023composing}. None treats direct teammate
nominations as a protected input.
Recent work has explored hybrid approaches that combine preference matching
with constraint satisfaction. \citet{holmberg2019metaheuristics} formulates student group
formation as a mathematical optimization problem and reports simulated
annealing and tabu search implementations at course scale. \citet{yannibelli2018evolutionary} combine a genetic algorithm
with simulated annealing for role-balanced teams, and
\citet{albashabsheh2025_lexicographic} order faculty constraints
before student preferences in capstone allocation, where preferences
concern projects rather than teammates.

In educational-technology venues, grouping studies include
swarm-intelligence metaheuristics for the student grouping problem in
e-learning systems~\citep{alhunitah2016swarm}, particle swarm
optimization over learner characteristics for differentiated
instruction~\citep{zervoudakis2020pso}, and group formation coupled with
peer assessment in MOOCs~\citep{haddadi2018peer}. These optimize
composition against instructor- or model-defined criteria; none takes
direct teammate nominations as a protected input.

Within operational research practice, integer programming has been
used to form balanced, cohesive student teams from instructor-specified
criteria at Indiana University's Kelley School
\citep{cutshall2007indiana} and to build MBA study groups at Toronto's
Rotman School \citep{krass2006rotman}; efficiency and fairness criteria
for student--project assignment have been formalized in the
optimization literature \citep{rezaeinia2022fairness}; and the mixed-integer formulation with solver and metaheuristic
comparisons of \citet{holmberg2019metaheuristics} is the closest
methodological benchmark for the present problem. A recent multi-method review finds little evidence linking
fairness definitions to groupwork outcomes and argues that
deterministic, explainable allocation supports accountability
\citep{forshaw2025fairness}. None of these takes direct teammate
nominations as a protected first-stage input.

\subsection{Research gap}\label{sec:gap}

Among the systems and deployment studies reviewed, we found no
large-cohort implementation that treats direct teammate nominations as a
protected first-stage input while subsequently optimizing academic
balance and gender-isolation criteria. Existing tools either configure composition without protecting
nomination matches or hand students the criteria without balance
guarantees \citep{kifle_bonner_2023}. Within OR
practice, deployed integer-programming systems optimize
instructor-defined composition
\citep{cutshall2007indiana,krass2006rotman}, but the comparative value
of protecting direct teammate nominations in a separate stage has not
been quantified.

In a single-stage objective, preference and fairness terms compete
directly: an accepted move may improve balance by breaking a nomination
match. The method developed here instead fixes small preference-based units
before optimizing the pairing of those units.

\subsection{Contribution}\label{sec:contribution}

We address this gap with a two-phase procedure. Phase~1 uses weighted graph
heuristics to form fixed units of two or three students from direct
teammate nominations. Phase~2 uses simulated annealing to pair the units
into teams, seeking academic balance while penalizing gender isolation and
departures from the target team size. The study makes three contributions as an account of operational research practice:

\begin{enumerate}
  \item \textbf{Practice-informed problem formulation.} A
    translation of conflicting administrator and student requirements
    into a two-stage decision process: protect direct teammate
    nominations as fixed units, then optimize the pairing of those
    units under balance, isolation, and size penalties.
  \item \textbf{Deployment and operational evaluation.}
    Implementation with 238 students in two cohorts, evaluated
    against a six-year archive of 82 historical grouping instances
    (1{,}538 teams) and Monte Carlo random-assignment nulls, with
    administrative benefits and limitations documented.
  \item \textbf{Reproducible implementation lessons.} Open software
    archived with a version DOI, an integerised conditional
    optimisation benchmark for validation, a documented data-pipeline
    failure and its correction, deterministic tie-breaking, and a
    deployment checklist (Section~\ref{sec:checklist}).
\end{enumerate}

Not all team-related dissatisfaction stems from formation; free-riding
during the project also contributes \citep{oakley2004turning}.
Contribution-based assessment arrangements operated alongside the
formation algorithm and are reported separately. The present study is
limited to team formation.

\section{Methods}\label{sec:methods}

\subsection{Overview of the two-phase approach}

The procedure handles preference matching and cohort-level balance in
separate stages. This local-then-global structure resembles hierarchical
optimization in materials processing and also connects the allocation
exercise to content taught in the host module. Each cohort is processed
independently:
\begin{itemize}
  \item \textbf{Phase 1}: form fixed units that prioritize teammate
    nominations;
  \item \textbf{Phase 2}: pair those units by simulated annealing to
    improve academic balance and penalize isolation or invalid team sizes.
\end{itemize}

\subsection{Data preparation and constraint definition}

\subsubsection{Input data}
The algorithm takes a student roster (identifier, registry gender,
programme), GPAs on a 4.0 scale, and up to two teammate nominations
per student. The historical archive records prior attainment on the
institutional 0--100 mark scale (student-level range 31--96), so all
cross-era comparisons use scale-free metrics
(Section~\ref{sec:evaluation}).

\subsubsection{Allocation criteria}
Let $S = \{s_1, s_2, \ldots, s_n\}$ represent the set of students in a
cohort, where $n \approx 120$. Each student $s_i$ has attributes
\begin{align}
  s_i &= \{id_i,\; gpa_i,\; gender_i,\; prefs_i\},\\
  prefs_i &= \{p_{i1}, p_{i2}\} \subseteq S \setminus \{s_i\}.
\end{align}

The allocation targets the following outcomes:
\begin{itemize}
  \item team size: $|T_j| = 6 \pm 1$ for each team $T_j$;
  \item gender isolation: for each recorded gender category $c$,
    $\forall T_j$, $|T_j^c| \neq 1$;
  \item academic balance: minimize the variance of team mean GPAs.
\end{itemize}
These are implemented as weighted penalties in Phase~2 rather than as
hard constraints (Eq.~\ref{eq:cost}); achieved team sizes and isolation
counts are therefore checked after optimization.

\subsection{Phase 1: preference-driven unit formation}

Phase~1 uses the nomination graph to form fixed units, giving priority to
mutual and densely connected nominations.

\subsubsection{Preference graph construction}
We construct a directed graph $G = (V, E)$ where vertices $V = S$ (all
students), edges $E = \{(s_i, s_j) \,|\, s_j \in prefs_i\}$.
Mutual preference pairs are identified as
\begin{equation}
  M = \{(s_i, s_j) \,|\, (s_i, s_j) \in E \land (s_j, s_i) \in E\}.
\end{equation}
From $E$ we form the symmetrized nomination graph $H$ on $V$: an
unordered edge $\{s_i, s_j\}$ is present whenever $(s_i, s_j) \in E$
or $(s_j, s_i) \in E$, with weight 2 for mutual nominations and 1 for
one-way nominations. Phase~1 operates on $H$.

\subsubsection{Hierarchical assignment strategy}
Students are assigned to triads following a strict preference hierarchy.

Units are formed by a strict four-level priority cascade: (1)
3-cliques of $H$ (triads in which every pair is linked by a
nomination in either direction), ranked by total edge weight and
assigned greedily subject to disjointness---a triangle need not be
fully mutual, and the deployed run broke ties by enumeration order
where v1.3 uses ascending identifiers; (2) unassigned mutual pairs
completed by the best-scoring third member; (3) one-way nominators
grouped with an available nominee; (4) GPA-balanced triads formed
from the sorted remainder, with any final one or two students forming
a remainder unit. The formal scoring rules and the full pipeline
listing are given in Supplementary Material (Section~S5).

\subsection{Phase 2: simulated annealing for balanced pairing}

Phase 2 employs simulated annealing, a metaheuristic inspired by
metallurgical annealing in which materials are heated and slowly cooled to
reach minimum-energy states \citep{kirkpatrick1983optimization}. The
connection to materials science provided a direct pedagogical link in our
polymer physics context (Section~\ref{sec:pedagogy}).

\subsubsection{Initial configuration}
Units are sorted by mean GPA and mirror-paired (unit $i$ with unit
$n_u{-}1{-}i$); an odd middle unit would remain unpaired and incur
the lone-unit penalty. With 119 students, Phase~1 yields 39 triads
plus one two-student remainder ($n_u = 40$), so pairing gives 19
six-person teams and one five-person team; the size and lone-unit
penalties are retained for general cohort sizes.

\subsubsection{Objective (as implemented)}
The deployed objective is a score with a 100-point baseline from which
weighted penalties are subtracted:
\begin{equation}
\mathrm{score}(\Pi) = 100
 - w_v\,\mathrm{Var}(\bar g)
 - w_r\,\bigl(\max_j \bar g_j - \min_j \bar g_j\bigr)
 - w_i\,N_{\mathrm{iso}}
 - w_s \sum_{j \in \mathcal{P}_2} \bigl|\,|T_j| - 6\,\bigr|
 - w_u\,N_{\mathrm{lone}},
\label{eq:cost}
\end{equation}
where $\mathrm{Var}(\bar g)$ is the population variance of team mean
GPAs, the range term penalizes the spread between the best- and
worst-off teams, $N_{\mathrm{iso}}$ counts teams containing a
gender-isolated individual, and $\mathcal{P}_2$ indexes teams formed by
pairing two Phase~1 units. The size term penalizes those paired teams for
deviation from six members; an unpaired unit instead contributes to
$N_{\mathrm{lone}}$. The weights
$w_v = 1000$, $w_r = 50$, $w_i = 20$, $w_s = 5$, $w_u = 10$ were set
empirically before deployment and are subjected to a sensitivity sweep
in Section~\ref{sec:robustness}. Gender isolation and team size thus
enter as weighted penalties rather than hard constraints; their
satisfaction is verified on the deployed output
(Section~\ref{sec:gender_results}). Higher scores are better; the
attainable maximum is below 100 whenever a five-person team exists.
Substituting the deployed configurations into
Eq.~\ref{eq:cost} reproduces the reported final scores (87.4
and 75.6) exactly.

\subsubsection{Annealing schedule}\label{sec:annealing}
The temperature schedule follows exponential decay, $T(t) = T_0 \alpha^t$,
with $T_0 = 100$ and $\alpha = 0.995$. The run terminates when
$T < 0.01$, after approximately 1{,}840 iterations; the
10{,}000-iteration upper bound is therefore not reached.

\subsubsection{Neighbor generation and acceptance}
A neighbour swaps one Phase-1 unit between two teams; acceptance
follows the Metropolis criterion on the score difference, accepting
improvements always and deteriorations with probability
$\exp(\Delta s / T)$.

\subsection{Implementation and tool}\label{sec:implementation}

The reference implementation (v1.3) is written in Python
(\texttt{pandas}, \texttt{networkx}, \texttt{numpy},
\texttt{openpyxl}). Clique enumeration has an $O(n^3)$ worst-case
bound and Phase-2 evaluation is $O(km)$ over $k$ proposals and $m$
units; each 119-student cohort ran in 3--5 seconds on an Intel i7
computer with 16\,GB RAM, and runtime beyond this cohort size was
not benchmarked. The release validates nominated students against the
roster and handles asymmetric nominations (approximately 64\% of
nominations) explicitly; the deployment reader used a less robust
free-text parser whose effect is quantified in
Section~\ref{sec:phase1_results}.

\subsubsection{Use of generative AI tools}
OpenAI ChatGPT (GPT-5.6 Thinking) and Anthropic Claude (Claude Fable 5)
were used for language editing, consistency checks, \LaTeX{} formatting,
and assistance in developing the analysis and verification scripts. The
authors independently reviewed the text, reran and verified the reported
analyses, confirmed the originality and accuracy of the content, and accept
full responsibility for the manuscript, code, references, and results.

\subsection{Evaluation design}\label{sec:evaluation}

We evaluate the deployment on four outcome families: preference
satisfaction, academic balance, gender composition, and the institutional
complaint record. Because the historical archive and the deployment cohort
record academic performance on different scales
(Section~\ref{sec:baseline}), all cross-era academic-balance comparisons use
a scale-free \emph{balance ratio}, and within-cohort claims are benchmarked
against explicit Monte Carlo null models.

\subsubsection{Metrics}
We evaluate four outcome families. \emph{Preference satisfaction}:
the share of submitters receiving at least one requested teammate,
reported on the all-submitter and machine-read bases
(Section~\ref{sec:phase1_results}). \emph{Balance ratio} $\rho$:
the between-team standard deviation of team mean GPAs divided by its
expectation under uniformly random assignment into the same team-size
structure,
\begin{equation}
  \rho = \frac{\mathrm{SD}_{\mathrm{obs}}(\bar{g}_1, \ldots,
  \bar{g}_K)}{\mathbb{E}_{\mathrm{null}}\!\left[\mathrm{SD}(
  \bar{g}_1, \ldots, \bar{g}_K)\right]},
  \label{eq:rho}
\end{equation}
scale-free across mark scales and eras; reported SDs use the sample
convention while Eq.~\ref{eq:cost} uses population variance, matching
the deployed code. \emph{Gender composition}: teams containing a
gender-isolated individual, and single-gender teams. \emph{Formal
complaints}: written objections through official channels; per-student
counts were not archived, so historical prevalence is reported as
documented recurrence rather than as a rate (full definitions:
Supplementary Material, Section~S6).

\subsubsection{Monte Carlo null model}\label{sec:nullmodels}
A random-assignment null draws 10{,}000 uniformly random
partitions of each cohort into the realized team-size multiset,
yielding null distributions of the between-team SD and the isolated-
team count, with one-sided empirical $p$-values
$(1 + \#\{\mathrm{null} \leq \mathrm{observed}\})/(N+1)$. The
same null, run per historical instance on its own data, supplies the
denominator of Eq.~\ref{eq:rho}. Gender isolation and team size enter the optimizer as weighted
penalties (Eq.~\ref{eq:cost}); we verify both on the deployed
output and report the null expectation for context. Because the
annealing is stochastic, Section~\ref{sec:robustness} reports outcome
distributions across 50 optimizer seeds per cohort and a one-at-a-time
weight-sensitivity sweep.

\subsection{Benchmark and comparator design}\label{sec:comparators}

\emph{Integerised conditional benchmark.} Holding the deployed Phase-1 units
fixed, the pairing stage is a perfect matching of 40 units over 780
candidate pairs whose every objective term is a function of the chosen
pairs. Branching on the two-student remainder's partner makes the sum
of selected team means constant within each branch, so each of the 39 branches is a linear matching model. Formally,
with binary $x_{uv}$ selecting the pairing of units $u$ and $v$,
degree constraints $\sum_{v} x_{uv} = 1$ for every unit, and the
remainder unit's partner $p$ branched over the 39 triads, each branch
minimises $\sum_{uv} c^{(p)}_{uv} x_{uv} + 50\,(\mathrm{mx} -
\mathrm{mn})$, where $c^{(p)}_{uv} = 50\,(\mu_{uv} - \bar g_p)^2 +
\mathrm{pen}_{uv}$ collects the variance contribution about the
branch-fixed grand mean $\bar g_p$ together with the isolation and
size penalties, and $\mathrm{mx}$, $\mathrm{mn}$ bound the selected
team means. CP-SAT (OR-Tools 9.15) proved global optimality for this
integerised Phase-2 objective, conditional on the fixed Phase-1
units; every branch returned OPTIMAL within a 30-second per-branch
limit on laptop-class hardware. Team means and edge costs are
integerised at $10^{-6}$ resolution, and the reported optima are
exact-objective re-evaluations of the solver-optimal matchings. No
2-exchange improves either solution under the exact objective (a
local check), and a corrected surrogate matching (per-branch grand
mean, per-edge variance weight $1000/20$, penalties included, range
omitted by design) selects the identical pairing in both cohorts.
Baselines are the mirror initialization, a greedy matching on true
edge costs, and 1{,}000 random unit matchings.

\emph{Single-stage comparator.} A student-level simulated
annealing over the same team-size structure optimizes the deployed
objective plus $w_p$ per satisfied nominator ($w_p \in
\{0,1,3,10,30,100\}$; 50 seeds per setting) at the deployed budget
($\approx$1{,}838 proposals) and, for $w_p \in \{0,10,100\}$, a
tenfold budget ($\alpha=0.9995$) to demonstrate convergence.
Outcomes: between-team SD, isolation, satisfied nominators, double
nominators receiving both nominees, and stability (retention of the
deployed matches; mean pairwise Jaccard of matched-edge sets across
seeds). The both-nominee subanalysis used 20 seeds for the
single-stage ($w_p=10$) condition at each budget; two-phase results
used 50 seeds.

\subsection{Historical baseline}\label{sec:baseline}

To situate the deployment against prior practice, we analyzed historical
team assignment data from two related laboratory-based modules (UK levels 4 and 5, delivered in the second and third years of the
four-year joint program) spanning six academic years
(2019--2024). The archive
comprises 82 distinct grouping instances, 1{,}538 teams, and 9{,}534
student--team assignments.

\subsubsection{Academic balance}
Between-team variance of team mean marks varied widely across historical
instances: mean $9.74 \pm 14.32$, median $4.05$, range $0.31$--$86.21$, in
squared 0--100 mark units (2023 instances averaged 31.25); such
magnitudes are impossible on the 4.0 GPA scale, which is why
cross-era comparison requires a scale-free metric. Computing the
balance ratio $\rho$ (Eq.~\ref{eq:rho}) per historical instance from its
own student-level data gives mean $\rho = 0.87 \pm 0.44$ (median 0.85,
range 0.31--2.55): manual assignment was on average modestly more balanced
than chance, but with wide spread, and 18 of the 82 instances were less
balanced than random assignment.

\subsubsection{Gender composition}
Gender balance presented a persistent challenge in these cohorts, where
male students typically comprise 75--85\% of enrollment. Across all
historical groupings, 548 of 1{,}538 teams (35.6\%) contained only a single
gender; a further 79 teams (5.1\%) contained exactly one isolated minority
member; and instances averaged 6.7 single-gender teams each.

\subsubsection{Qualitative complaint record}
Institutional records document persistent student dissatisfaction across multiple years (Student--Staff Liaison Committee minutes, 2021--2025; available from the corresponding author): entries from 2021--22, 2023--24, and 2024--25 record requests to change the grouping method, to form teams freely, and complaints about unequal contribution (verbatim entries in Supplementary Material, Section~S3).
Figure~\ref{fig:historical_comparison} summarizes the historical balance
and gender-composition distributions against the deployed algorithm.

\begin{figure}[t]
\centering
\includegraphics[width=0.95\textwidth]{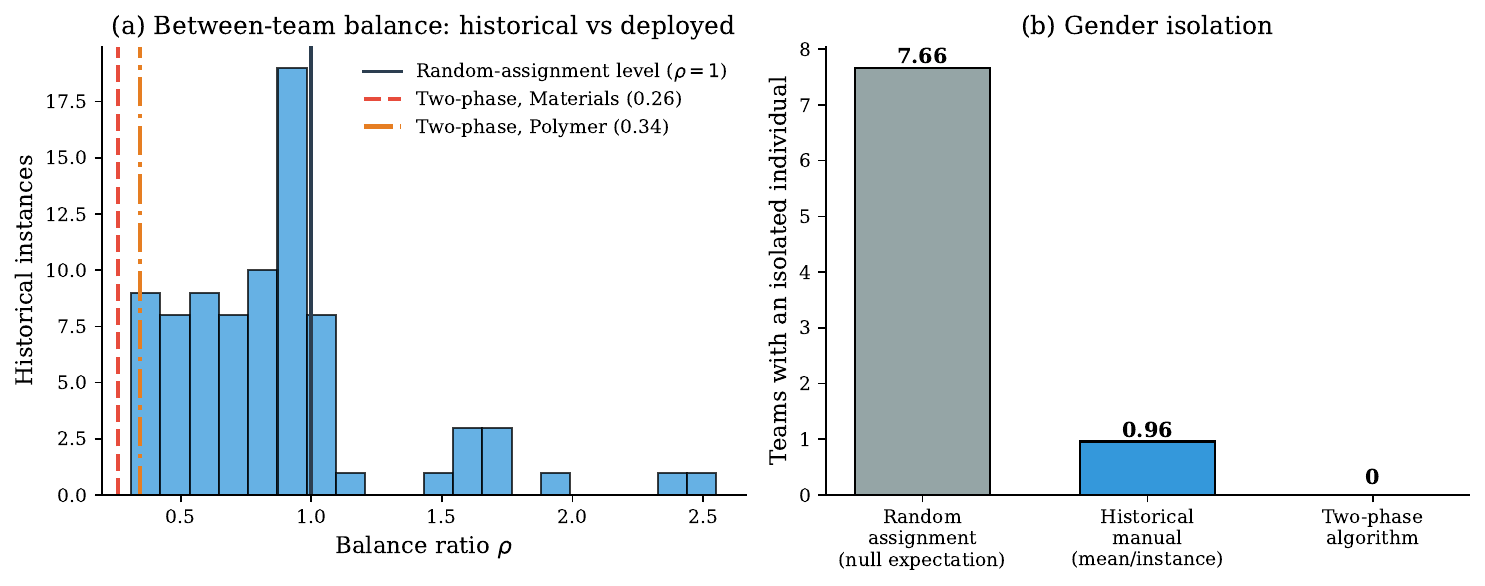}
\caption{Team formation quality against the six-year historical archive.
(a) Distribution of the per-instance balance ratio $\rho$
(Eq.~\ref{eq:rho}) across 82 historical grouping instances (2019--2024),
with the deployed cohorts and the random-assignment level ($\rho = 1$)
marked. (b) Teams containing a gender-isolated individual: the expectation
under random assignment (mean of the two deployment cohorts), the
historical manual mean per instance, and the deployed algorithm.
Single-gender teams (547 of 548 all-male in the archive) are not shown:
under the no-isolation criterion and the observed female share, some
all-male teams are unavoidable and are not treated as problematic when no
individual is isolated. The random benchmark uses the deployment cohorts
and is shown for context rather than as a like-for-like historical
comparison.}
\label{fig:historical_comparison}
\end{figure}

\section{Results}\label{sec:results}

\subsection{Overall performance}

The two-phase algorithm was deployed across 238 students (119 Materials,
119 Polymer) in a physical properties of polymers course.
Table~\ref{tab:overall_metrics} summarizes outcomes against the historical
archive.

\begin{table}[H]
\caption{Team formation outcomes: historical archive vs.\ two-phase
deployment.}
\label{tab:overall_metrics}
\centering
\small
\begin{threeparttable}
\begin{tabularx}{\textwidth}{
  >{\raggedright\arraybackslash}p{4.6cm}
  >{\raggedright\arraybackslash}X
  >{\raggedright\arraybackslash}p{3.5cm}}
\toprule
Metric & Historical archive\tnote{a} & Two-phase deployment \\
\midrule
Formal grouping-related complaints & documented recurrently in the reviewed 2021--2025 records\tnote{b} &
  \textbf{0} (238 students) \\
Preference satisfaction ($\geq$1 requested teammate) & not collected &
  \textbf{80.4\%} of submitters\tnote{c} \\
Balance ratio $\rho$ (1 = random-assignment level) &
  $0.87 \pm 0.44$ & \textbf{0.26} / \textbf{0.34} \\
Teams containing a gender-isolated individual & 79/1{,}538 (5.1\%) &
  \textbf{0}/40 \\
Single-gender teams & 548/1{,}538 (35.6\%) &
  19/40 (47.5\%)\tnote{d} \\
\bottomrule
\end{tabularx}
\begin{tablenotes}
\footnotesize
\item[a] 82 grouping instances across six academic years
  (2019--2024); 547 of the 548 single-gender teams were all-male, one
  all-female.
\item[b] Student--Staff Liaison Committee minutes, 2021--2025; per-student
  complaint counts were not archived (see Section~\ref{sec:evaluation}).
  Available from the corresponding author on request.
\item[c] 115 of 143 students whose forms named at least one identifiable
  cohort member: 76.6\% (82/107) Materials, 91.7\% (33/36) Polymer. Among
  the 96 nomination sets machine-read in the deployed run: 90.6\% (87/96);
  the bases differ because of a parsing divergence described in
  Section~\ref{sec:phase1_results}.
\item[d] An expected consequence of the no-isolation criterion: pairing
  female students concentrates them, increasing the number of all-male
  teams while ensuring no individual is isolated.
\end{tablenotes}
\end{threeparttable}
\end{table}

No formal grouping-related complaint was recorded in either cohort.
Complaint rates are rarely reported in the algorithmic team-formation
literature, which more often evaluates satisfaction or team viability
\citep{layton2010catme,hastings2020lift}; a like-for-like external
comparison is therefore unavailable. The institutional minutes document
grouping concerns in more than one year between 2021 and 2025, whereas no
such concern was recorded for the deployment cohort. This descriptive
contrast is treated as a proxy outcome and is not attributed to the
algorithm alone (Section~\ref{sec:limitations}).

\subsection{Phase 1: preference-based unit formation}\label{sec:phase1_results}

Submission differed sharply between cohorts: 107 of 119 Materials
students (89.9\%) and 36 of 119 Polymer students (30.3\%) named at
least one identifiable cohort member, of which the deployed run
machine-read 61 and 35. Post hoc verification found two parser
faults: the tokenizer split on commas only, missing 46
space-separated Materials submissions, and the reader skipped
Polymer's first data row. Satisfaction is therefore reported on both
the all-submitter and machine-read bases
(Section~\ref{sec:limitations}).

Phase 1 operated on the machine-read nomination sets and identified
strong social structure in Materials: the nomination graph contained 31
mutual pairs and yielded 18 nomination triangles---triads in which every
pair is linked by at least one nomination, five of them fully
mutual---placing 45.4\% of the cohort in a triad with two teammates
they nominated or were nominated by. The full reconstructed unit inventory is given in
Supplementary Material (Section~S2), and Section~\ref{sec:robustness}
recovers the deployed units themselves.
The parsed Polymer nominations formed no complete triangles, and
satisfaction there came through one-way preference matching. The high triangle incidence (clustering 0.46 all students, 0.64
among nominators, against 0.01--0.02 at random) suggests strong
pre-existing friendship networks.

Among machine-read submitters, 88.5\% (54/61, Materials) and 94.3\%
(33/35, Polymer) received at least one requested teammate: 90.6\% pooled
(87/96). Re-parsing every submission post hoc, the corresponding figures
across all submitters are 76.6\% (82/107) and 91.7\% (33/36): 80.4\%
pooled (115/143). Twenty-eight of the 46 Materials students whose forms were not
machine-read still received a requested teammate, 22 via a
groupmate's machine-read reciprocal nomination. Under the
random-assignment null with the realized team structures, a
submitter would receive at least one nominee in roughly 8\% of cases
(7.3--8.2\% across cohorts and bases), so the observed rates represent
roughly a tenfold increase over chance.

\subsection{Phase 2: balance and composition optimization}

\begin{figure}[!htbp]
\centering
\includegraphics[width=0.8\textwidth]{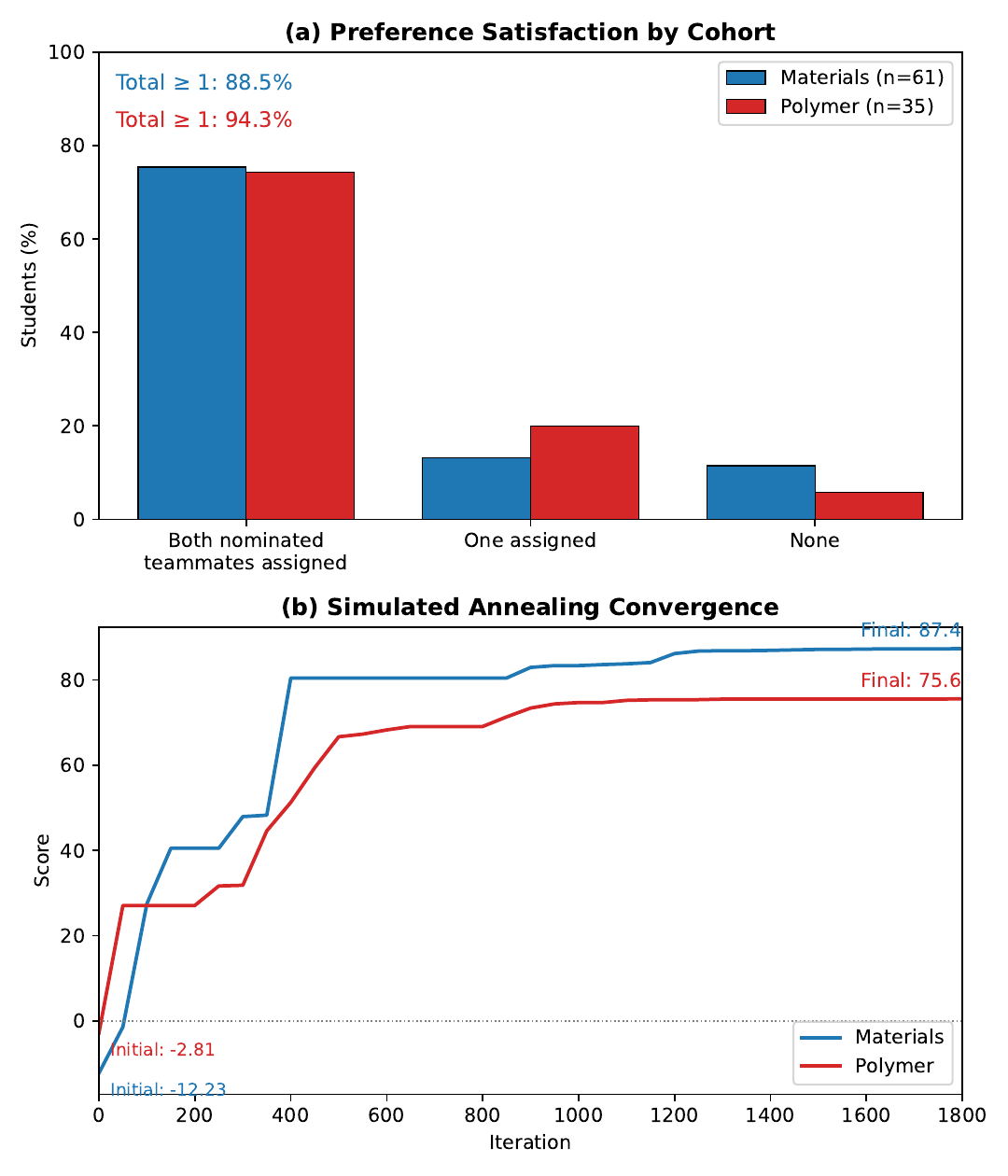}
\caption{Phase 1 and Phase 2 performance. (a) Preference satisfaction by cohort. (b) Simulated annealing convergence (score $= 100$ minus weighted penalties, Eq.~\ref{eq:cost}): initial scores of $-12.23$ (Materials)
and $-2.81$ (Polymer) improved to 87.4 and 75.6, respectively. Panel (a) uses the machine-read submitter basis (Section~\ref{sec:phase1_results}).}
\label{fig:phase_results}
\end{figure}

The annealing converged within the $\approx$1{,}840-iteration
schedule (Figure~\ref{fig:phase_results}): snake-draft
initialization scores of $-12.23$ and $-2.81$ improved to $87.4$ and
$75.6$. On the recovered
deployed units (Section~\ref{sec:robustness}), the mirror-paired
initialization already produced low between-team variance
($1.1 \times 10^{-3}$ and $4.5 \times 10^{-3}$, population
convention); the large negative initial scores instead reflect gender
isolation in the initial configurations (five and four isolated teams),
so the annealing gain consisted chiefly of eliminating isolation while
preserving balance.

\subsection{Academic balance}\label{sec:balance_results}\label{sec:gpa_results}

\begin{figure}[!htbp]
\centering
\includegraphics[width=0.95\textwidth]{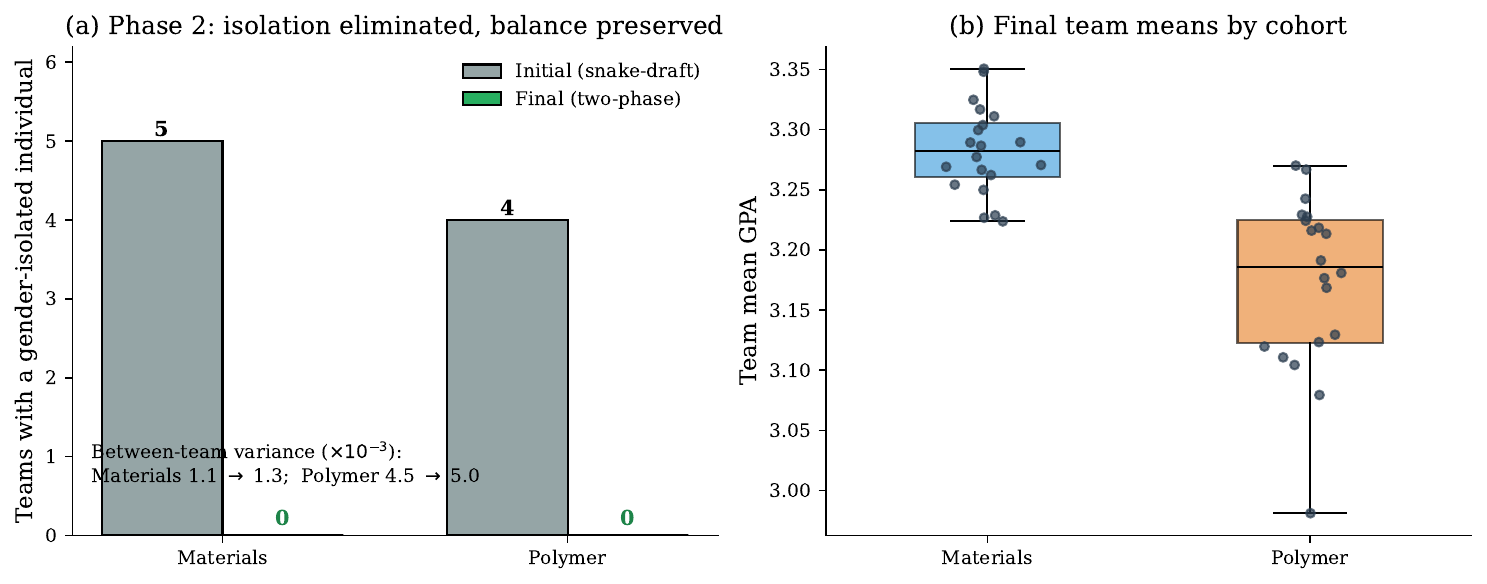}
\caption{Between-team academic balance in the deployed cohorts.
(a) Effect of Phase~2 on the snake-draft initialization, computed on the
recovered deployed units (Section~\ref{sec:robustness}): the annealing
eliminated gender isolation (five and four isolated teams initially;
none finally) while between-team variance of team mean GPAs was
essentially preserved (Materials $1.1 \to 1.3$, Polymer $4.5 \to 5.0$,
$\times 10^{-3}$, population convention). (b) Distribution of final
team mean GPAs by cohort; boxes span the interquartile range, whiskers
the full range, individual teams overlaid.}
\label{fig:gpa_distribution}
\end{figure}

Team means spanned 0.127 GPA points (3.224--3.351; SD 0.037) in
Materials and 0.289 (2.981--3.270; SD 0.072) in Polymer
(Figure~\ref{fig:gpa_distribution}); one Polymer team sat 2.7
standard deviations below the cohort's team-mean average.

\begin{figure}[!htbp]
\centering
\includegraphics[width=0.95\textwidth]{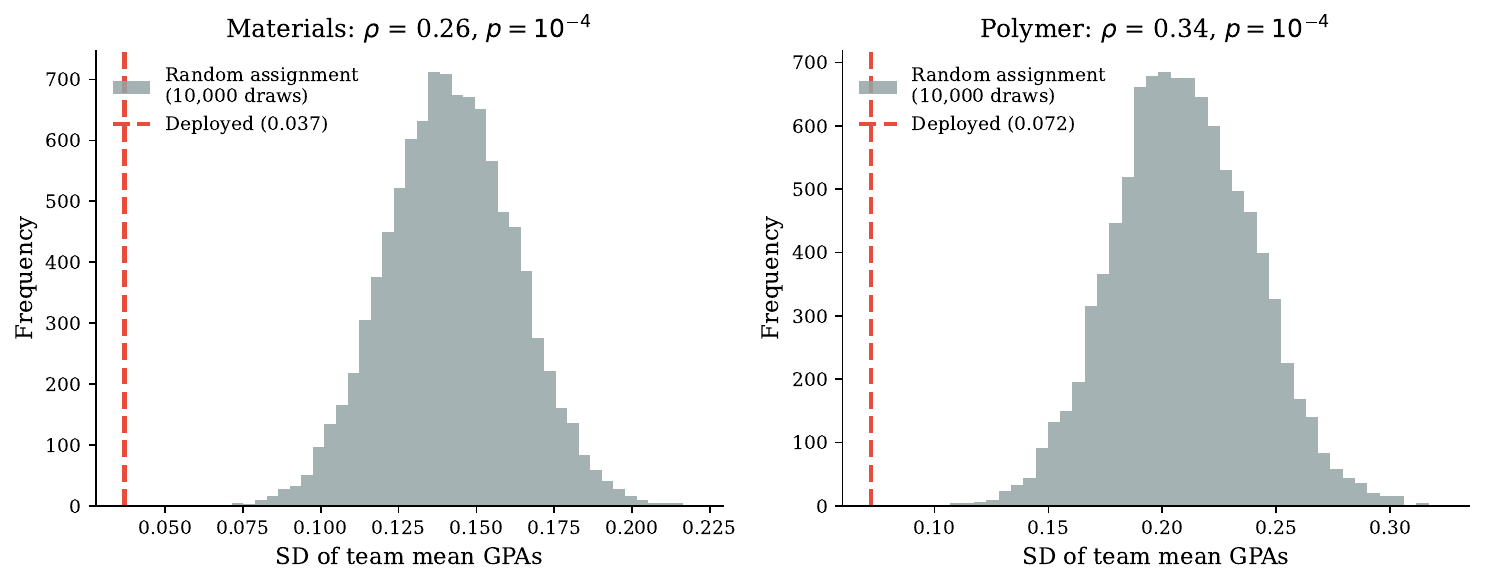}
\caption{Random-assignment null distributions of the between-team standard
deviation of team mean GPAs (10{,}000 draws per cohort), with the deployed
configurations marked. No draw reached the deployed spread in either
cohort (empirical $p = 1.0 \times 10^{-4}$).}
\label{fig:mcnull}
\end{figure}

Against the random-assignment null (Section~\ref{sec:nullmodels}), the
deployed configurations achieved balance ratios $\rho$ of 0.26 (Materials;
observed between-team SD 0.037 against a null expectation of 0.142) and
0.34 (Polymer; 0.072 against 0.210). In 10{,}000 random partitions per
cohort, no draw matched the deployed spread (empirical
$p = 1.0 \times 10^{-4}$ in both cohorts), and the single best random draw
remained 1.5--1.9 times wider than the deployed configuration. On the same
scale-free metric, historical manual assignment averaged $\rho = 0.87 \pm
0.44$: the Materials deployment lies below all 82 historical instances,
and the Polymer deployment below 79 of the 82.
Figure~\ref{fig:mcnull} shows the null distributions.

\subsection{Gender composition}\label{sec:gender_results}

No team in either cohort contained a gender-isolated individual:
whenever female students were present, at least two were placed
together (gender-balanced composition in every team is arithmetically
impossible at 75--85\% male enrollment). The 40 teams comprised 21
mixed-gender and 19 all-male teams. By contrast, the historical archive averaged 0.96 teams with an
isolated individual per grouping instance (79 of 1{,}538 teams, 5.1\%),
although historical cohort compositions and team structures differed
from those of the deployment cohorts. Under the random-assignment null,
the expected number of teams containing an isolated individual was 7.1
(Materials) and 8.2 (Polymer), and a zero-isolation partition arose at
most once in 10{,}000 random draws per cohort. The deployed algorithm
reduced the observed count to zero.

\subsection{Computational performance}

Each 119-student cohort was processed in 3--5 seconds on an Intel i7
computer with 16\,GB RAM. The nomination graphs were sparse (average degree 1.4 in Materials
and 1.0 in Polymer), and Phase~2 terminated after approximately 1{,}840 iterations
(Section~\ref{sec:annealing}).

\subsection{Robustness across optimizer seeds and weights}
\label{sec:robustness}

Phase~1 is deterministic for a fixed nomination file and tie-breaking
rule, so stochasticity enters only through Phase~2. The released
robustness script evaluates two Phase~1 unit inventories.

First, the \emph{deployed} units themselves, recovered directly
from the deployed output workbook (members are listed in unit order);
the recovered units reproduce the deployed initial and final scores
to export rounding (details: Supplementary Material, Section~S2). Re-running the deployed annealing with the
same neighbor generator, schedule, and objective, we tested 50 seeds per
cohort. Final scores were 86.6--87.7 for Materials (median 87.5) and
73.0--75.9 for Polymer (median 75.0). The respective between-team SD
ranges were 0.035--0.045 and 0.070--0.077; balance-ratio ranges were
$\rho=0.248$--$0.315$ and $0.335$--$0.366$. The deployed
configurations ($\rho = 0.26$ and 0.34) lie within these
distributions, at the 40th and 34th percentiles, respectively. No seed produced a gender-isolated individual in either cohort
(0 of 100 runs), and the team-size structure (19 six-person teams, one
five-person team) is fixed by the unit
inventory in these cohorts.

A one-at-a-time weight sweep (50 seeds per setting) on the deployed
units left outcomes stable across most order-of-magnitude
perturbations; the exceptions were in the lower-submission Polymer
cohort, where cutting the isolation weight tenfold produced isolated
teams in 5 of 50 runs and inflating the variance weight tenfold in 1
of 50 (full grid: Supplementary Material, Table~S1). Materials produced no
isolated team in any of its 350 baseline and weight-sweep runs.

Second, \emph{reconstructed} units: re-deriving Phase~1 from the
archived preference forms reproduces the deployed run's logged counts
but agrees with the deployed partition on only 28 and 26 of the 40
units, and re-runs on the reconstructed units do not reach the
deployed Materials balance (details in Supplementary Material, Section~S2);
the released implementation (v1.3) therefore uses identifier-based
tie-breaking.

\subsection{Exact benchmark and single-stage comparison}
\label{sec:benchmark_results}

\emph{Integerised conditional optimum.} The integerised
conditional optima of the pairing stage are 87.73 (Materials) and
75.88 (Polymer): the deployed solutions lie 0.37 and 0.27 points below
optimal on the 100-point objective, and the best of the 50 re-seeded
annealing runs attained 87.73 and 75.86. No 2-exchange improves either optimum under the exact objective (a
local check), and the corrected surrogate matching selects the
identical pairing in both cohorts. The optimal solutions have
between-team SDs of 0.035 and 0.070 with zero isolation. For scale, the greedy-cost baseline scores 45.8 and 23.7, the
mirror initialization $-12.1$ and $-2.5$, and 1{,}000 random unit
matchings have medians near $-100$ (best 38.0 and 3.6): the baselines
performed substantially worse, and the deployed heuristic solved the
pairing to near-optimality.

\begin{figure}[!htbp]
\centering
\includegraphics[width=0.98\textwidth]{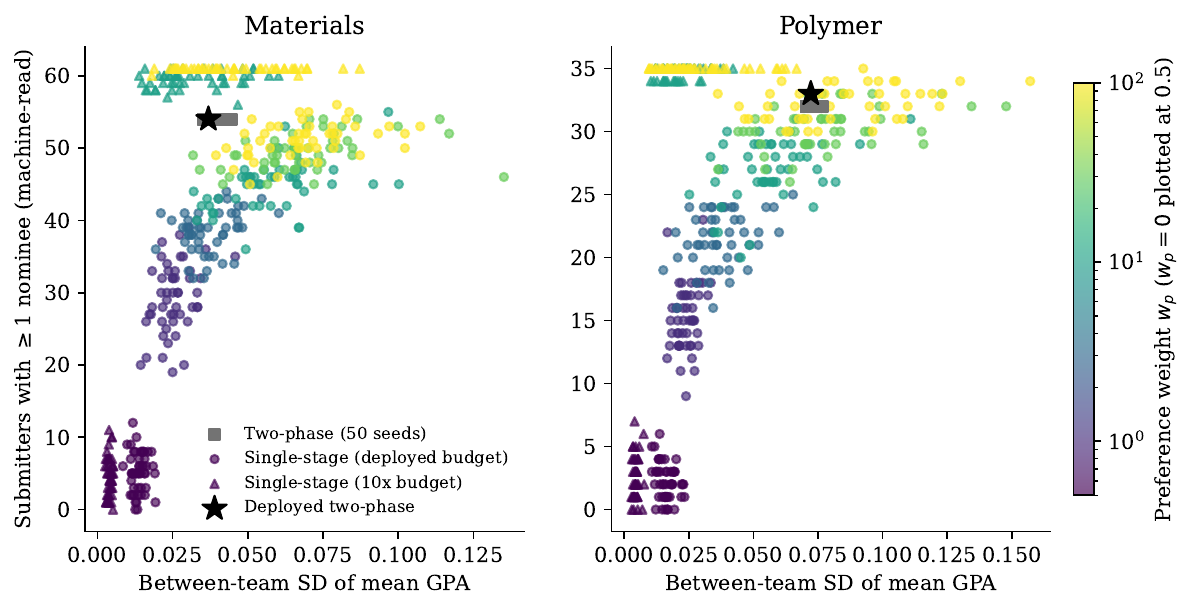}
\caption{Satisfaction--balance outcomes across methods. Grey
squares: 50 two-phase re-runs on the deployed units; black star: the
deployment; circles and triangles: single-stage runs at the deployed
and tenfold budgets, coloured by preference weight $w_p$. At matched budget no
single-stage setting's median reaches the deployed operating point
(one Polymer run matched it; see text); at tenfold budget single-stage
exceeds it on both plotted axes, saturating the satisfaction axis (the
horizontal band at the top of each panel), while rarely delivering both nominees (a separate 20-seed $w_p{=}10$ subanalysis; see text) and, at high $w_p$ and matched budget, frequently violating the isolation criterion.}
\label{fig:pareto}
\end{figure}

\emph{Budget-matched single-stage comparison.} At the deployed search budget, no single-stage setting's median
reaches the deployed operating point (Figure~\ref{fig:pareto}); a
single run of the 600 (Polymer, $w_p=100$) matched it, with 33
satisfied at SD 0.072 and no isolation. Median satisfied nominators peak at 51 of 61 and 32 of 35 at
$w_p=100$---below the Materials two-phase re-run median of 54 and
level with Polymer's 32 (deployed 33)---at SDs of 0.069 and 0.079
against the two-phase 0.035--0.045 and 0.070--0.077, and with
isolated teams in 18--20 and 23--37 of 50 runs at $w_p \geq 30$: a
large preference reward in a single objective overwhelms the 20-point
isolation penalty, the competition the decomposition was designed to
avoid.

\emph{Extended-budget comparison.} With a tenfold budget and
$w_p=10$, the single-stage optimizer surpasses the deployment on
aggregate satisfaction and balance simultaneously: median 60 of 61 and
35 of 35 nominators satisfied at SDs of 0.027 and 0.021 with zero
isolation. At this budget the reward saturates: the medians sit at the
attainable ceiling of the machine-read nomination sets (all 61 and all
35 submitters at $w_p=100$; 60 of 61 at $w_p=10$ in Materials), which
is why the long-budget points form a horizontal band at the top of
each panel of Figure~\ref{fig:pareto}, and further increases in $w_p$
trade only balance. Aggregate $\geq$1 satisfaction, however, conceals match quality.
The single-stage comparator retained a median of 16 of 52 Materials
students and 0 of 26 Polymer students with both nominated teammates
per run, whereas Phase~1 protected both nominees for 44 of 52 Materials
double nominators and all 26 in Polymer, and the deployed Phase~2
raised the Materials final-team count to 46: one shared teammate
satisfies the single-stage reward, so the optimizer stops there.
Matched pairs also remain seed-dependent (mean pairwise Jaccard 0.81
and 0.46); protected Phase-1 matches are invariant by construction,
and in these data the complete set of satisfied nomination edges was
unchanged across the 50 Phase-2 re-runs. The comparison quantifies the trade: fixing units limits balance
(best observed Materials SD of 0.048 across the 100 reconstructed-unit
runs, against 0.021--0.027 attainable without fixed units) in exchange
for both-nominee match quality and invariant protected matches,
without calibrating a preference-reward weight against balance.

\section{Discussion}\label{sec:discussion}

\subsection{Benefits of the two-phase decomposition}

Because Phase~1 units are atomic during annealing, Phase~2 cannot
break a match already secured inside a unit.
The trade-off is explicit and now quantified: fixing the units
limits attainable balance (best observed Materials SD of 0.048 across the 100 reconstructed-unit runs, against 0.021--0.027 reached by unconstrained single-stage search at
extended budget), and yields both-nominee match quality (44 of 52 protected in
Phase~1 and 46 in the deployed Materials teams; 26 of 26 in Polymer,
against single-stage medians of 16 and 0), Phase-1 matches that are
invariant by construction (the satisfied-edge set was unchanged across
all 50 re-runs), and stable behaviour at deployment budgets without
calibrating a preference-reward weight
(Section~\ref{sec:benchmark_results}). A single-stage formulation
with nomination matches imposed as hard constraints could combine the
strengths of both designs and is a natural next step. This differs from a single objective in which
preference and fairness weights compete throughout the search. The
local-then-global structure also mirrors hierarchical optimization in
materials processing and was used as a worked course example
(Section~\ref{sec:pedagogy}).

The 18 nomination triangles in the Materials cohort (clustering 0.46 against $\approx 0.01$ for a random graph of equal
density) indicate a
strongly transitive nomination network in this cohort. Clique detection
may be less useful where nomination density is lower.

\subsection{Relation to existing systems}

Table~\ref{tab:comparison} positions the approach against
established systems on capability rather than outcome, since
evaluation settings differ.

\begin{table}[H]
\caption{Capability comparison of team formation approaches.}
\label{tab:comparison}
\centering
\small
\begin{tabularx}{\textwidth}{
  >{\raggedright\arraybackslash}p{2.8cm}
  >{\raggedright\arraybackslash}X
  >{\raggedright\arraybackslash}X
  >{\raggedright\arraybackslash}p{3.0cm}}
\toprule
Approach & Preference input & Composition constraints & Reported setting \\
\midrule
Random assignment & none & balance only in expectation & any \\
Self-selection & full (self-organized) & none; remainder problem & any \\
CATME \citep{layton2010catme} & instructor-configured criteria & yes &
  large cohorts \\
Team Anneal \citep{uq_teamanneal} & none & yes, incl.\ isolation rules &
  large cohorts \\
LIFT \citep{hastings2020lift} & collective criteria voting & configurable &
  289 students (experiment) \\
\textbf{Two-phase (ours)} & \textbf{direct teammate nominations} &
  \textbf{isolation and size verified; GPA balance optimized} & \textbf{238 students (deployment)} \\
\bottomrule
\end{tabularx}
\end{table}

Here, students provide person-specific nominations that are
protected before the balance search begins; the systems above
configure allocation criteria or let students choose them.

\subsection{Optimizer choice in deployment and in reuse}

Simulated annealing was the method actually deployed; the exact
benchmark was developed retrospectively to audit that deployment. The
audit shows the deployed solutions were near-optimal (within 0.37 and
0.27 points), so the deployed outcomes were within 0.4 points of the benchmark, but for future use at this
scale the exact matching formulation is preferable when its solver dependency is acceptable: exact matching is certifiable and admits hard composition constraints. The annealing implementation remains relevant
as a dependency-free, easily extensible alternative whose behaviour at
deployment budgets is now characterized, with incidental teaching value in the host module (Section~\ref{sec:pedagogy}).

\subsection{Practical implications}

According to the module coordinator, manual allocation had
required at least two hours and sometimes half a day per cohort; the
optimization ran in 3--5 seconds, excluding preference collection,
roster validation, and review of the proposed teams. No
grouping-related complaint was recorded through the institutional
channels reviewed for the deployment year, although no student survey
was administered, so perceptions of the process were not measured
(Section~\ref{sec:limitations}). Simulated annealing is taught in
the host module, and the instructor used the allocation as a worked
example of the algorithm\label{sec:pedagogy}; this study did not
measure learning from that example, so its pedagogical value should
be tested rather than inferred.

\subsection{A deployment checklist}\label{sec:checklist}

The deployment's failure modes translate into a short checklist for
anyone operating allocation tools on live student data:
\begin{itemize}
  \item collect nominations against structured student identifiers,
    not free-text names;
  \item validate every input record against the roster before
    optimization, and report accepted and rejected records to the
    operator;
  \item make all tie-breaking deterministic so that runs are
    reproducible;
  \item independently reproduce the output from the archived inputs
    before release;
  \item have a human review the proposed teams before publication to
    students.
\end{itemize}
Had the first two items been in place, the tokenizer fault of
Section~\ref{sec:phase1_results} could not have silently dropped 46
submissions.

\subsection{Limitations and boundary conditions}\label{sec:limitations}

The measured outcomes are administrative; preference satisfaction and
the complaint record are proxies, and the absence of recorded
complaints is evidence only of no expressed dissatisfaction through
official channels, not of team effectiveness. Because complaint
ascertainment was retrospective, and because team-based assessment
involves arrangements beyond formation (a new contribution-based
mark-adjustment scheme was introduced in the same year), the
complaint contrast is descriptive and not attributed to the algorithm
alone; team performance and learning outcomes were not measured. The
integerised optimum is conditional on the deployed Phase-1 units, the
global joint optimum was not computed, and the single-stage
comparator probes that space over a finite weight grid and two
budgets with a reward targeting satisfied nominators, the reported
aggregate; rewarding both-nominee matches directly would re-create
triad protection.

The evaluation covers one institution, one deployment year, and one
transnational engineering programme, compared against a six-year
internal archive whose instances span two modules, a different mark
scale, and varying team structures; the per-instance $\rho$
computation absorbs this, but the records were not collected for
research. The registry gender field is binary, so the isolation
criterion reflects administrative categories only.

Free-text nomination capture is a real failure mode: the deployed
tokenizer missed 46 of 107 Materials submissions
(Section~\ref{sec:phase1_results}); v1.3 splits on commas and
whitespace, and the browser interface reads validated structured
columns. When fewer students submit (36 of 119 in Polymer), fewer
matches can be protected, so future deployments should announce the
process early and report both response and satisfaction rates.
Runtime was measured only for the two 119-student cohorts;
larger-cohort performance should be benchmarked before use.

\subsection{Future directions}

The next evaluation should collect direct measures of team functioning,
student perceptions, and learning outcomes. Further work could also test
skill-based matching, adjustment after peer feedback, and repeated
allocation across modules. Multi-site replication is needed to determine
which results depend on the local cohort and nomination network.

\section{Conclusion}\label{sec:conclusion}

The two-phase procedure preserved nomination matches before optimizing
the pairing of fixed units. In two 119-student cohorts, 80.4\% of all
valid submitters received a requested teammate, between-team GPA spread
was 26\% and 34\% of the random-assignment expectation, and none of the
40 teams contained a gender-isolated student. The deployed pairings were proven within 0.37 and 0.27 points of the integerised conditional optimum; re-seeded runs reproduced the balance and
isolation results, and the weight sweep showed isolation reappearing
when its penalty is weakened. A single-stage comparator exceeded the
deployment on aggregate satisfaction and balance only at a tenfold
search budget, while rarely delivering both nominated teammates; Phase-1-protected matches are invariant by construction.

The deployment supports the feasibility of the method at this cohort size,
but not a causal claim about complaints, team performance, or learning.
Those outcomes require prospective measurement and replication beyond one
site. 
The open-source implementation and verification scripts are provided to
support such replication.

\section*{Acknowledgments}
The authors thank Dr Folashade Akinmolayan Taiwo for discussions about
development and dissemination of the toolkit.

\section*{Statements and Declarations}

\subsection*{Funding}
This work was supported by the CREME Scholarship Support Fund, School of
Engineering and Materials Science, Queen Mary University of London, for
the project ``Algorithmic Team Formation Toolkit: From Zero Complaints to
Universal Implementation.''

\subsection*{Competing interests}
The authors have no relevant financial or non-financial interests to
disclose.

\subsection*{Ethics approval}
Queen Mary University of London's Research Ethics Facilitator, acting on
behalf of the Queen Mary Research Ethics Committee, screened a Research
Ethics Application covering this work and determined that no further
ethical review was required because the project is an evaluation rather
than a research project (reference QME26.2676, 28 July 2026). The
evaluation used routinely collected educational records from 2019--2025,
including rosters, academic averages, teammate preferences, and committee
minutes. The records were processed and reported in anonymized,
cohort-level aggregate form.

\subsection*{Data availability}
Code and de-identified derived data are available at
\url{https://github.com/sundanny89-debug/two-phase-team-formation}
and are archived in Zenodo as Version 1.3, DOI:
\href{https://doi.org/10.5281/zenodo.21792712}{10.5281/zenodo.21792712}.
Individual-level roster, attainment, and nomination records are not
publicly available because they are protected student educational
records and public release was not authorized; any access is subject
to Queen Mary University of London data-governance requirements.

\subsection*{Author contributions}
Y.S.: conceptualization, methodology, software, formal analysis,
investigation, data curation (deployment), visualization, funding
acquisition, project administration, writing---original draft. X.D.:
resources (six-year historical grouping archive), data curation (archive),
methodology (prior formation practice and feedback), writing---review and
editing. D.G.P.: conceptualization (problem identification), resources
(deployment module and historical complaint records), investigation
(co-implementation), writing---review and editing.

\subsection*{Supplementary information}
Supplementary Material contains the robustness grid, reconstructed Phase-1
unit results, liaison-committee entries, released-output inventory,
formal Phase-1 rules and pipeline listing, and evaluation-metric
definitions.

\bibliographystyle{unsrtnat}
\bibliography{references}

\end{document}